\documentclass[pra,twocolumn]{revtex4-1}
\usepackage{xcolor}
\usepackage{graphicx}
\usepackage{amssymb, amsmath, amsthm}
\usepackage{bbold}
\usepackage{dsfont}
\usepackage[percent]{overpic}

\usepackage[colorlinks]{hyperref}

\usepackage[authormarkup=none]{changes}
\definechangesauthor[name={a}, color=orange]{a} 
\definechangesauthor[name={s}, color=blue]{c} 

\def \diracspacing {0.7pt}
\newcommand{\bra}[1]{\langle #1 \hspace{\diracspacing} |} 
\newcommand{\ket}[1]{| \hspace{\diracspacing} #1 \rangle} 
\newcommand{\braket}[2]{\langle #1 \hspace{\diracspacing} | \hspace{\diracspacing} #2 \rangle} 

\newcommand{\dg}{\dagger}
\newcommand{\cD}{\mathcal{D}}
\newcommand{\cH}{\mathcal{H}}
\newcommand{\cO}{\mathcal{O}}
\newcommand{\id}{\hat{\mathds{1}}}
\newcommand{\rmd}{\mathrm{d}}

\begin{document}
	
\author{Alexandre Roulet}
\affiliation{Department of Physics, University of Basel, Klingelbergstrasse 82, CH-4056 Basel, Switzerland}

\author{Christoph Bruder}
\affiliation{Department of Physics, University of Basel, Klingelbergstrasse 82, CH-4056 Basel, Switzerland}

\title{Synchronizing the Smallest Possible System}
\begin{abstract}
  We investigate the minimal Hilbert-space dimension for a system to
  be synchronized. We first show that qubits cannot be synchronized 
  due to the lack of a limit cycle. Moving to larger spin values, we
  demonstrate that a single spin 1 can be phase-locked to a weak
  external signal of similar frequency and exhibits all the standard
  features of the theory of synchronization. Our findings rely on the
  Husimi Q representation based on spin coherent states which we
  propose as a tool to obtain a phase portrait.
\end{abstract}
\maketitle
\emph{Introduction.--} The universal concept of synchronization applies to an impressively large variety of apparently unrelated systems. This ranges from the pendulum clocks of Huygens to heart beats, including light pulses emitted by fireflies as well as an applauding crowd~\cite{pikovsky01}. The hallmark of this phenomena is the possibility to adjust the phase of a self-sustained oscillator by perturbing it with an external periodic signal~\cite{adler46}, or coupling it to one or more self-sustained oscillators of similar frequencies~\cite{strogatz01}. They key component for this is the absence of a preferred phase of the oscillator, which can thus be locked while the limit cycle remains essentially intact.

The Van der Pol model~\cite{pikovsky01,vanDerPol26}, one of the workhorses for theoretical studies of this effect has recently
been formulated in the quantum framework, providing an excellent
platform for studying the impact of quantum effects onto
synchronization~\cite{tony13,walter14}. Building on the rapid
experimental progress~\cite{bagheri13,matheny14,zhang15,gilsantos17},
an exciting direction is to study how synchronization emerges in a
large network of such quantum oscillators~\cite{Ludwig2013}, similar
in spirit to the classical model proposed by
Kuramoto~\cite{kuramoto84}. However, when moving from a classical to a
quantum framework, one is inevitably confronted with the exponential
growth of the Hilbert space as opposed to the linear classical
scaling, which limits the tractable size of an arbitrary network of quantum oscillators.

In this Letter, we approach this issue
by addressing the fundamental question of the minimal elementary unit
which can be synchronized, using the Hilbert-space dimension as a
natural measure of size. While at first sight qubits and their associated Bloch
representation appear to be the ideal candidate for the smallest
possible system~\cite{zhirov08,ameri15,zhu15,galve17}, we show that they lack a
valid limit cycle and therefore do not fit into the standard paradigm of synchronization~\cite{pikovsky01}.
However, we find that the formalism of synchronization can be applied to the next smallest system: a single spin 1. 
To obtain a phase portrait applicable to spins in which we can
identify the phase variable that lies at the core of the
synchronization formalism, we choose spin coherent states, which undergo an oscillation similar to the closed curve followed by the coherent states of a harmonic oscillator in position-momentum space~\cite{radcliff71,arecchi72}. In particular, we demonstrate how the spin formulation of the Husimi Q representation can be used to visualize the limit cycle and provide signatures of synchronization.
\emph{Qubit.--} A natural attempt to find the smallest possible system
to synchronize is to consider a single qubit.
It comes equipped with a standard representation of the
2-dimensional Hilbert space $\cH_\text{qubit}$ known as the Bloch
sphere.
Any unitary applied to a qubit can be visualized as a rotation of some angle about some direction. As a consequence, for a generic Hamiltonian pointing in the direction $\hat{H}_\text{qubit}\propto \vec{n}\cdot\hat{\vec{\sigma}}$ where the vector of Pauli matrices is $\hat{\vec{\sigma}}=(\hat{\sigma}_x,\hat{\sigma}_y,\hat{\sigma}_z)$, any state different from the eigenstates $\ket{\pm \vec{n}}$ will rotate linearly in time about the direction set by the unit vector $\vec{n}$. This geometrical picture provides the necessary phase variable for addressing the question of synchronization, with the aim being to lock this oscillation in the Bloch sphere to an external signal.

To make contact with the standard paradigm of synchronization,
we first need to establish a valid limit cycle for the self-sustained oscillator. Specifically, adding loss and gain to the dynamics of the qubit, we look for a fixed point of the dissipative map that does not possess any phase preference. That the phase of the limit cycle needs to be free is a sine qua non condition which ensures that any perturbation neither grows nor decays, which is the essence of synchronization~\cite{pikovsky01}.

In the case of a qubit, any state belonging to the space $\cH_\text{qubit}$ can be written as $\hat{\rho}_\text{qubit}=(\id+\vec{m}\cdot\hat{\vec{\sigma}})/2$ where $|\vec{m}|\leq 1$ is maximized for pure states which lie on the surface of the Bloch sphere. This implies that the set of states invariant under rotations with respect to the axis $\vec{n}$ satisfy $\vec{m}=\lambda\vec{n}$ with $-1\leq\lambda\leq 1$. In other words, they correspond to probabilistic mixtures of the eigenstates $\ket{\pm \vec{n}}$, which are lying exactly on the rotation axis where the phase variable is not defined. Any attempt with qubits is thus bound to fail due to the absence of a phase-symmetric state that is different from the extremal eigenstates. When put into the context of a Van der Pol oscillator, this would mean being only able to stabilize the vacuum state, which remains at the center of the phase space and does not provide a valid limit cycle for synchronization. Indeed, the vacuum state is lacking self-sustained oscillations, which correspond to a stable closed curve in phase space with an associated phase variable.

\emph{Spin $S>1/2$.--} We now turn our attention to higher-spin
systems. Since the large-spin limit can be mapped to a harmonic
oscillator~\cite{primakoff40}, it is interesting to ask what
is the minimal value of $S$ required to observe the onset of
synchronization. To investigate this case, we first need to propose an
extension of the Bloch sphere representation to spins $S\geq 1/2$. The well-known harmonic-oscillator coherent states can be extended to spin systems and are obtained by rotating the extremal state~\cite{radcliff71,arecchi72},
\begin{equation}
	\ket{\theta,\phi}=\exp(-i \phi \hat{S}_z)\exp(-i\theta \hat{S}_y)\ket{S,S} ,
\end{equation}
where the spin operators are generators of the rotation group SO(3)
satisfying $[\hat{S}_j,\hat{S}_k]=i\epsilon_{jkl} \hat{S}_l$. Note
that this parametrization in terms of a pair of Euler angles $\theta$
and $\phi$ corresponds, in the case of a qubit, to the Bloch sphere representation. Moreover, we can easily show that the time evolution of
spin coherent states under the free Hamiltonian
\begin{equation}
  \hat{H}_0=\hbar\omega_0 \hat{S}_z
\end{equation}
is given by $e^{-i\hat{H}_0 t/\hbar}\ket{\theta,\phi}=\ket{\theta,\phi+\omega_0 t}$, which corresponds to an oscillation with frequency $\omega_0$ as desired. We thus have a family of states resembling a classical spin in a unit sphere pointing in the direction $\bra{\theta,\phi}\hat{\vec{S}}\ket{\theta,\phi} = \vec{n}_{\theta,\phi}$ where $\vec{n}_{\theta,\phi}=(\cos\phi\sin\theta,\sin\phi\sin\theta,\cos\theta)$, and precessing about the $z$ axis. This analogy can be made more rigorous by noting that these states form an overcomplete basis of the Hilbert space. Specifically, neighboring directions have an overlap given by $|\braket{\theta',\phi'}{\theta,\phi}|^2 = [(1+\vec{n}_{\theta,\phi}\cdot\vec{n}_{\theta',\phi'})/2]^{2S}$, which vanishes in the limit of a classical spin $S\to\infty$, and the completeness relation reads
\begin{equation}
	\int_0^\pi\!\rmd\theta\, \sin\theta \int_0^{2\pi}\!\rmd\phi\, \ket{\theta,\phi}\bra{\theta,\phi} =\frac{4\pi}{2S+1}\id .
\end{equation}

This motivates us to introduce the spin equivalent of the Husimi Q representation for discussing synchronization in spin systems, which is defined for any state $\hat{\rho}$ as~\cite{gilmore75}
\begin{equation}
	Q(\theta,\phi)=\frac{2S+1}{4\pi} \bra{\theta,\phi}\hat{\rho}\ket{\theta,\phi} .
\end{equation}
While it is normalized, this function should not be understood as a probability distribution,
since the three components of the spin operator $\hat{\vec{S}}$ do not
commute. Similar to other quasiprobability distributions such as the Wigner function, it is employed to provide a phase portrait, which in this case allows us to visualize any spin state in terms of the most ``classical'' states~\cite{husimi40}. Moreover, it is easily computed with the use of the Wigner D-matrix $D^S_{m',m}(\phi,\theta,0)=\bra{S,m'}e^{-i \phi \hat{S}_z}e^{-i\theta \hat{S}_y}\ket{S,m}$ which provides a representation of rotation operators in terms of the spin eigenbasis.

\begin{figure}
	\centering
	\begin{overpic}[width=0.5\columnwidth]{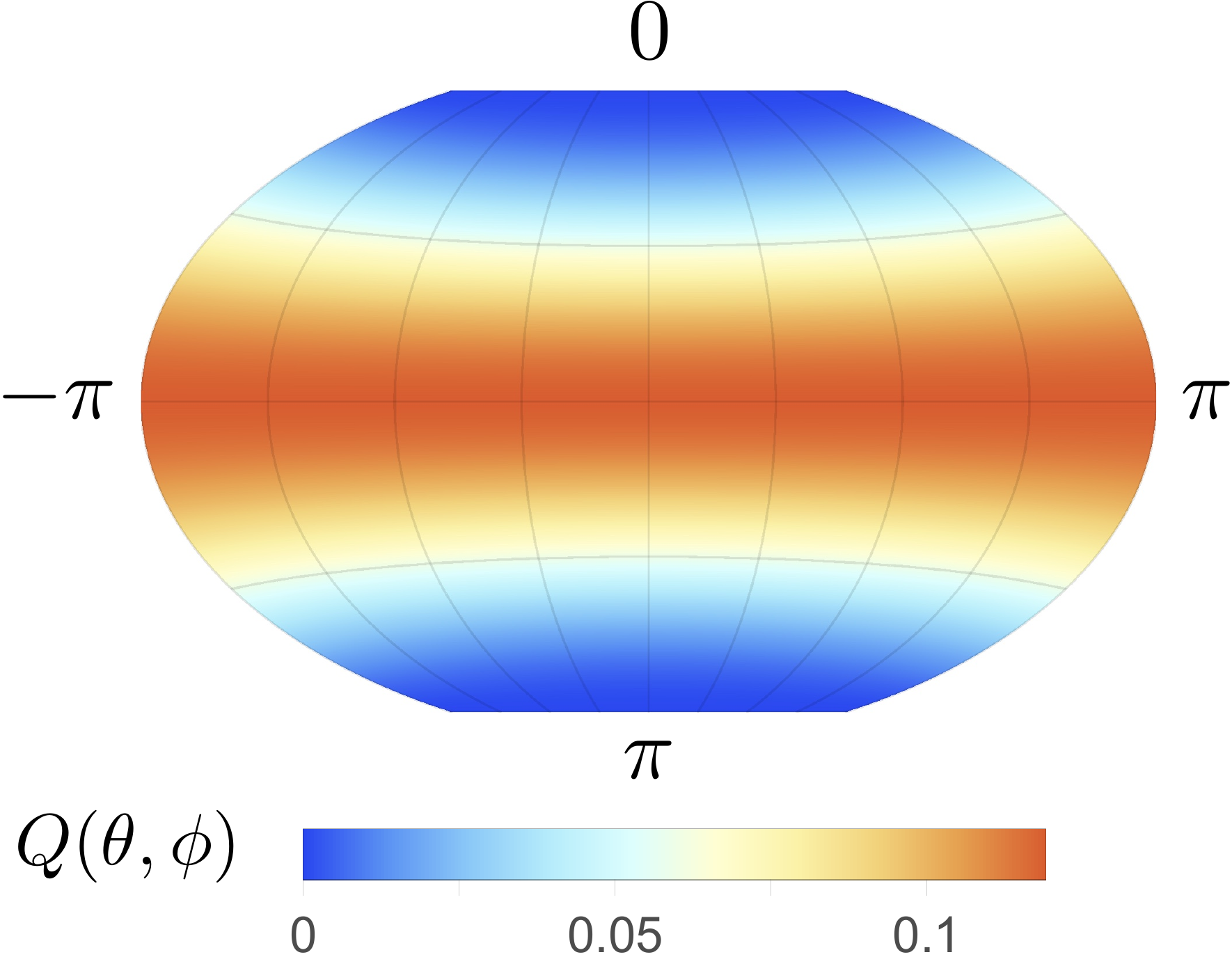}
		\put (0,70) {(a)}
	\end{overpic}\qquad
	\begin{overpic}[width=0.4\columnwidth]{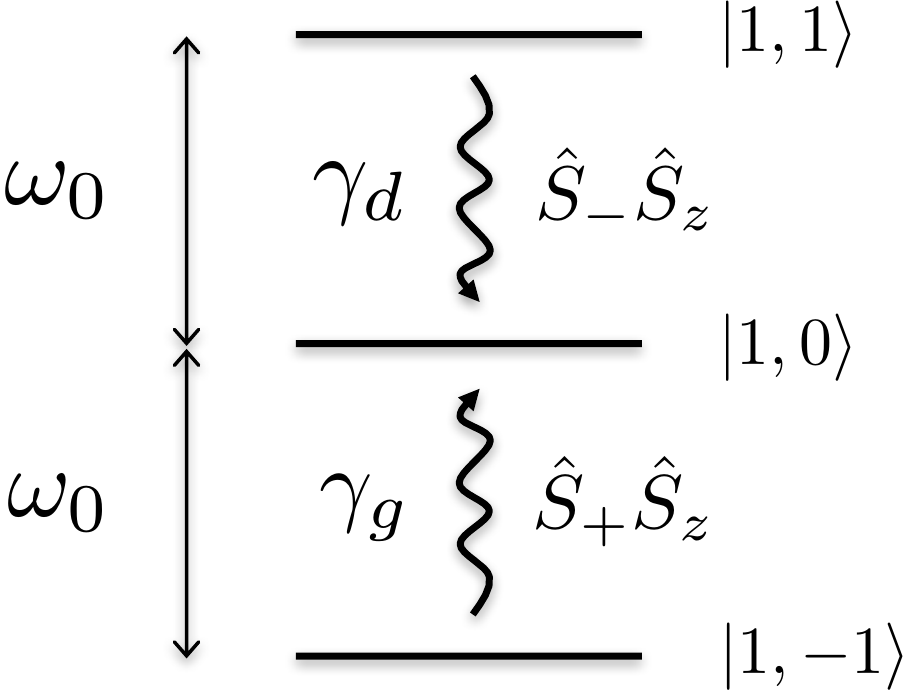}
		\put (0,87) {(b)}
	\end{overpic}
\caption{\label{fig:limitC} The limit cycle. (a) The Q function of the target state $\ket{1,0}$ represented using the Winkel tripel earth projection. The north and south poles correspond respectively to $\theta=0$ and $\theta=\pi$ while $\phi$ specifies the longitude. The $\phi$-symmetric distribution centered around the equator is reminiscent of the ring stabilized in the Wigner representation of a Van der Pol limit cycle~\cite{walter14}. (b) The spin-1 ladder, with the eigenstates separated by the natural frequency $\omega_0$. The action of the dissipators is to transfer the populations from the extremal states towards the target state.}
\end{figure}

\emph{Limit cycle.--} We will now apply this tool to study
synchronization in the context of a single spin 1. The key difference
with the qubit case is the existence of an additional phase-symmetric
state $\ket{1,0}$ that is not just a combination of the extremal eigenstates $\ket{1,\pm 1}$. Its Husimi Q representation $Q(\theta,\phi)=3 \sin^2(\theta)/8\pi$ is shown in Fig.\,\ref{fig:limitC}(a) using the Winkel tripel projection of a sphere~\cite{goldberg07}. It is independent of $\phi$ and centered around $\theta=\pi/2$, corresponding to coherent states on the equator precessing about the $z$ axis. The most straightforward approach to stabilize this target state is via the following master equation (in a frame rotating with $\hat{H}_0$) illustrated in Fig.\,\ref{fig:limitC}(b),
\begin{equation}\label{eq:masterDiss}
	\dot{\hat{\rho}}=\frac{\gamma_g}{2}\cD[\hat{S}_+\hat{S}_z ]\hat{\rho}+ \frac{\gamma_d}{2}\cD[\hat{S}_-\hat{S}_z ]\hat{\rho} ,
\end{equation}
where $\gamma_g$ and $\gamma_d$ are the respective gain and damping rates, and $\cD[\hat{\mathcal{O}}]\hat{\rho}=\hat{\mathcal{O}}\hat{\rho} \hat{\mathcal{O}}^\dg-\frac{1}{2}\left\{\hat{\mathcal{O}}^\dg\hat{\mathcal{O}},\hat{\rho}\right\}$ is the Lindblad superoperator~\cite{lindblad1976}. The dynamics generated by this dissipative map is not only attracting any state towards the equator, thereby stabilizing the energy, but does so without any phase preference. This is a consequence of the master equation being invariant under rotations about the $z$ axis. We have thus established a valid limit cycle, which we may now attempt to synchronize.
\begin{figure*}
	\centering
	\begin{overpic}[width=0.6\columnwidth]{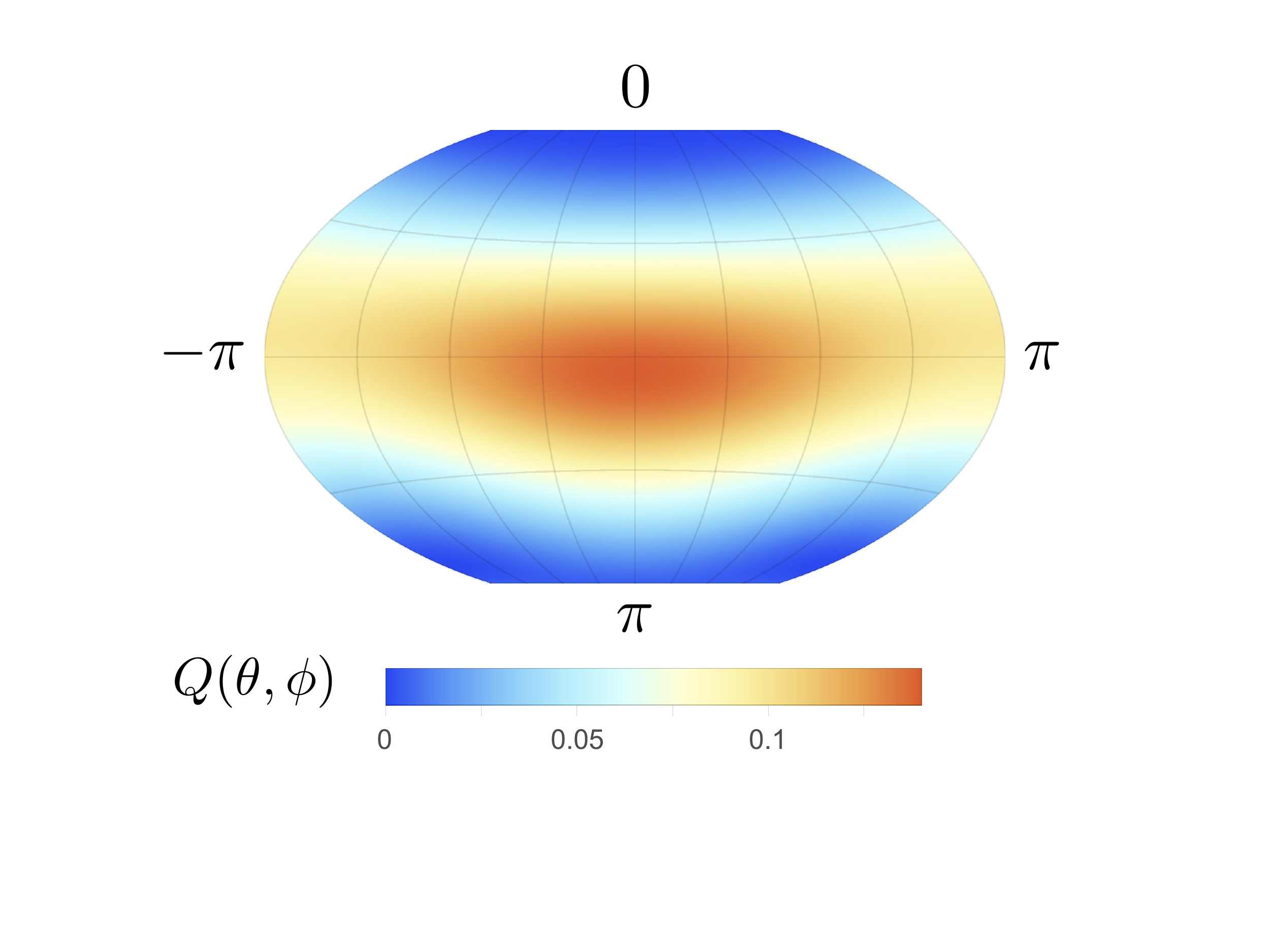}
		\put (0,70) {(a)}
	\end{overpic}\qquad
	\begin{overpic}[width=0.6\columnwidth]{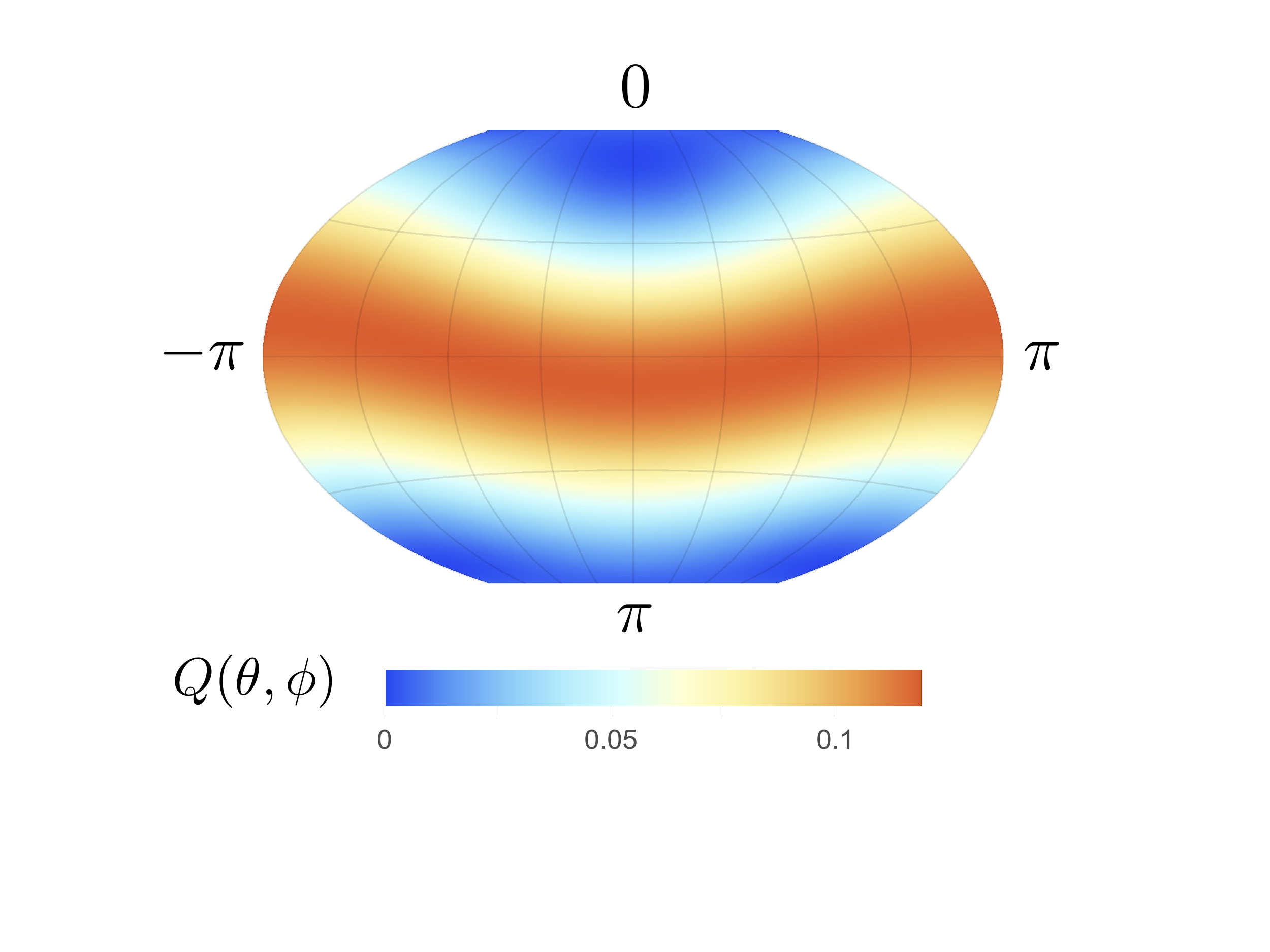}
		\put (0,70) {(b)}
	\end{overpic}\qquad
	\begin{overpic}[width=0.6\columnwidth]{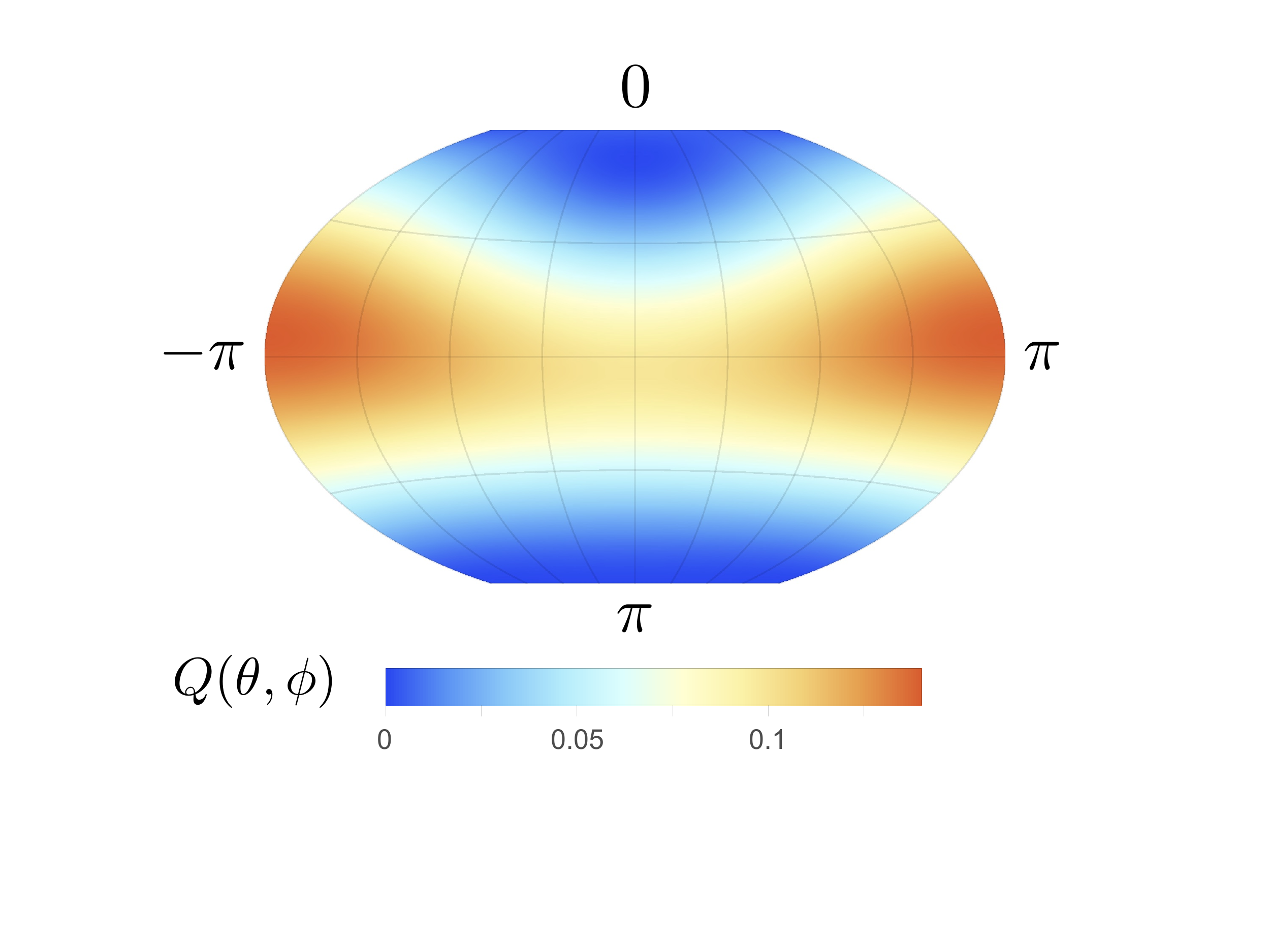}
		\put (0,70) {(c)}
	\end{overpic}
\caption{\label{fig:resPL} Phase locking to a resonant ($\Delta=0$)
  external signal represented by the steady-state Q function. The
  signal strength is chosen as $\varepsilon=0.1\gamma_\text{min}$ where $\gamma_\text{min}=\min(\gamma_g,\gamma_d)$,
  ensuring that we are in the weakly-perturbed regime. (a)
  $\gamma_g/\gamma_d=0.1$, $\gamma_g=10\epsilon$, the distribution is
  localized around $\phi=0$ while remaining approximately on the
  equator. (b) $\gamma_g/\gamma_d=1$, $\gamma_g=10\epsilon$, the limit
  cycle is slightly distorted with no visible signature of
  phase locking. (c) $\gamma_g/\gamma_d=10$, $\gamma_d=10\epsilon$,
  same as (a) but the distribution is now localized around $\phi=\pi$. Thus, phase locking is possible when one of the rates dominates.}
\end{figure*}
\emph{Synchronization.--} We model the external signal by a semi-classical drive of strength $\varepsilon$ at frequency $\omega$, described by the Hamiltonian in the rotating-wave approximation
\begin{equation}\label{eq:Hsig}
	\hat{H}_\text{signal}=i\hbar\frac{\varepsilon}{2}(e^{i\omega t}\hat{S}_- - e^{-i\omega t}\hat{S}_+ ) .
\end{equation}
Moving to a frame rotating with the drive frequency, the system is thus described by the master equation
\begin{equation}\label{eq:masterSynch}
	\dot{\hat{\rho}}=-i[\Delta \hat{S}_z+ \varepsilon \hat{S}_y,\hat{\rho}] + \frac{\gamma_g}{2}\cD[\hat{S}_+\hat{S}_z ]\hat{\rho}+ \frac{\gamma_d}{2}\cD[\hat{S}_-\hat{S}_z ]\hat{\rho} ,
\end{equation}
where $\Delta=\omega_0-\omega$ is the detuning between the frequency
of the autonomous oscillator and the drive. The steady state
can be obtained analytically, yet its expression
is rather uninformative as such. Instead, Fig.\,\ref{fig:resPL}
illustrates the power of the Q function to provide signatures of
synchronization. In particular, focusing first on the resonant
scenario we find that phase locking is achieved with equal or
opposite phases when one of the rates dominates.
This is one of the main results of our Letter. The $\pi$ phase difference between these two situations can be physically understood by noting that they are equivalent under time reversal with the relabelling $\ket{1,1}\leftrightarrow\ket{1,\!-\!1}$. Intriguingly, the
transition regime where the rates are comparable
shows no visible locking. 

To better understand these different synchronization regimes,
we
define the following synchronization measure,
\begin{equation}
	S(\phi)=\int_0^\pi\!\rmd\theta\, \sin\theta\, Q(\theta,\phi)-\frac{1}{2\pi} ,
\end{equation}
which vanishes everywhere in the absence of synchronization. On the
other hand, any non-uniform distribution of the phase will lead to a
positive maximum of this measure, indicating a locking to the corresponding phase~\cite{hush15,niels17}. We now take advantage of the small dimension of the Hilbert space and derive analytically the time evolution of this measure using the master equation~\eqref{eq:masterSynch}. Synchronization effects being of first order in the perturbation~\cite{pikovsky01}, we perform a first-order expansion in the signal strength, yielding
\begin{equation}\label{eq:Sdot}
	\dot{S}(\phi)=\frac{3\varepsilon}{16}(e^{-\gamma_g t/2}-e^{-\gamma_d t/2})\cos(\phi-\Delta t) +\cO\left([\frac{\epsilon}{\gamma_\text{min}}]^2\right) ,
\end{equation}
where $\gamma_\text{min}=\min(\gamma_g,\gamma_d)$. This expression
allows to compute the phase distribution at any time given the initial condition $S(\phi)=0\ \forall\phi$, corresponding to the phase-symmetric limit cycle (see Fig.\,\ref{fig:limitC}).

The first observation is that indeed, no synchronization is expected in the balanced case $\gamma_g=\gamma_d$ where $\dot{S}(\phi){\to} \cO\left([\frac{\epsilon}{\gamma_\text{min}}]^2\right)$. The weak signal is not able to adjust the phase, and increasing the power further will start to deform the limit cycle, thereby leaving the paradigm of synchronization. When one of the rates dominates, say $\gamma_d$ ($\gamma_g$), the exponential prefactor contributes positively (negatively) over a timescale of order $\gamma_g^{-1}$ ($\gamma_d^{-1}$). The dynamics is then that of synchronization to a phase reference that is linearly varying in time for non-zero detuning. When $|\Delta|\ll\gamma_\text{min}$, this phase reference is effectively constant and leads to significant in-phase (anti-phase) locking. This is the situation illustrated in Fig.\,\ref{fig:resPL}, for which the steady-state distribution is given from Eq.\,\eqref{eq:Sdot} by $S(\phi)\approx\frac{3\varepsilon}{8}(1/\gamma_g-1/\gamma_d)\cos(\phi)$. As the detuning is increased, the spin stays phase-locked to the shifted phase reference up to a regime $|\Delta|\gg\gamma_\text{min}$ where it cannot follow its fast rotation. The effect of synchronization is then averaged out, $\cos(\phi-\Delta t)\to 0$, as the signal is too off-resonant to adjust the phase.
\begin{figure*}
	\centering
	\begin{overpic}[width=0.59\columnwidth]{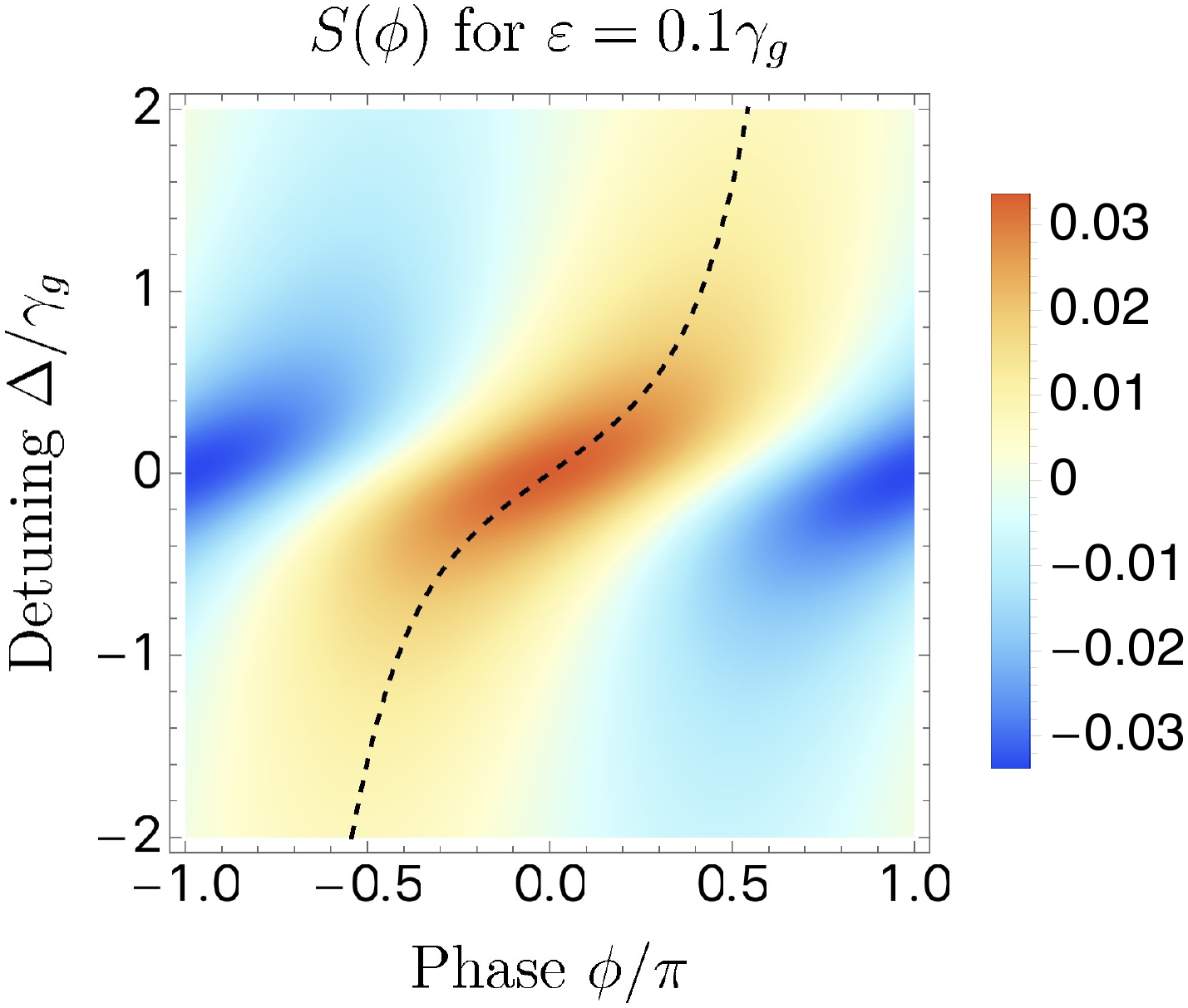}
		\put (0,85) {(a)}
	\end{overpic}\qquad\quad
	\begin{overpic}[width=0.62\columnwidth]{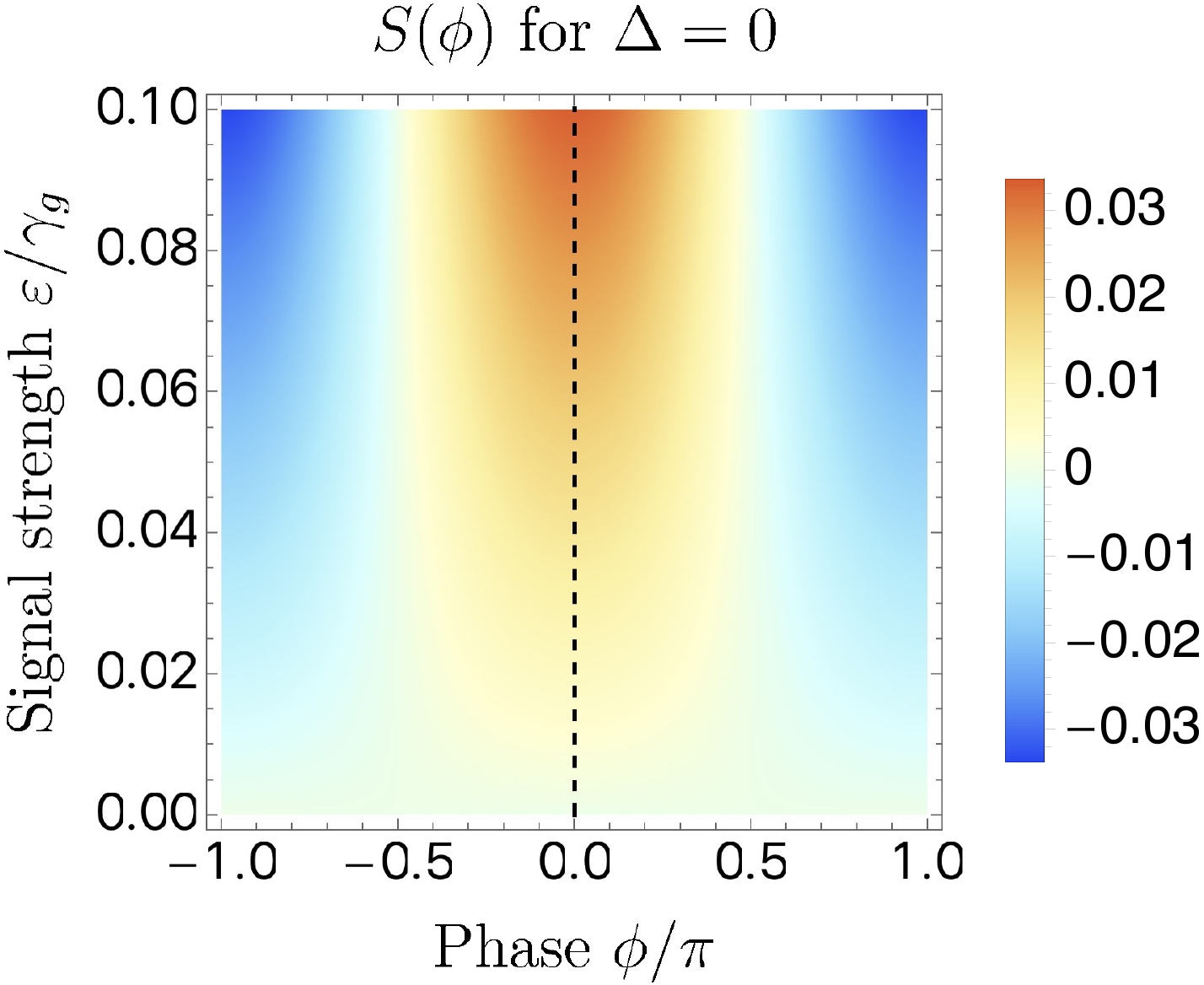}
		\put (0,82) {(b)}
	\end{overpic}\qquad\quad
	\begin{overpic}[width=0.605\columnwidth]{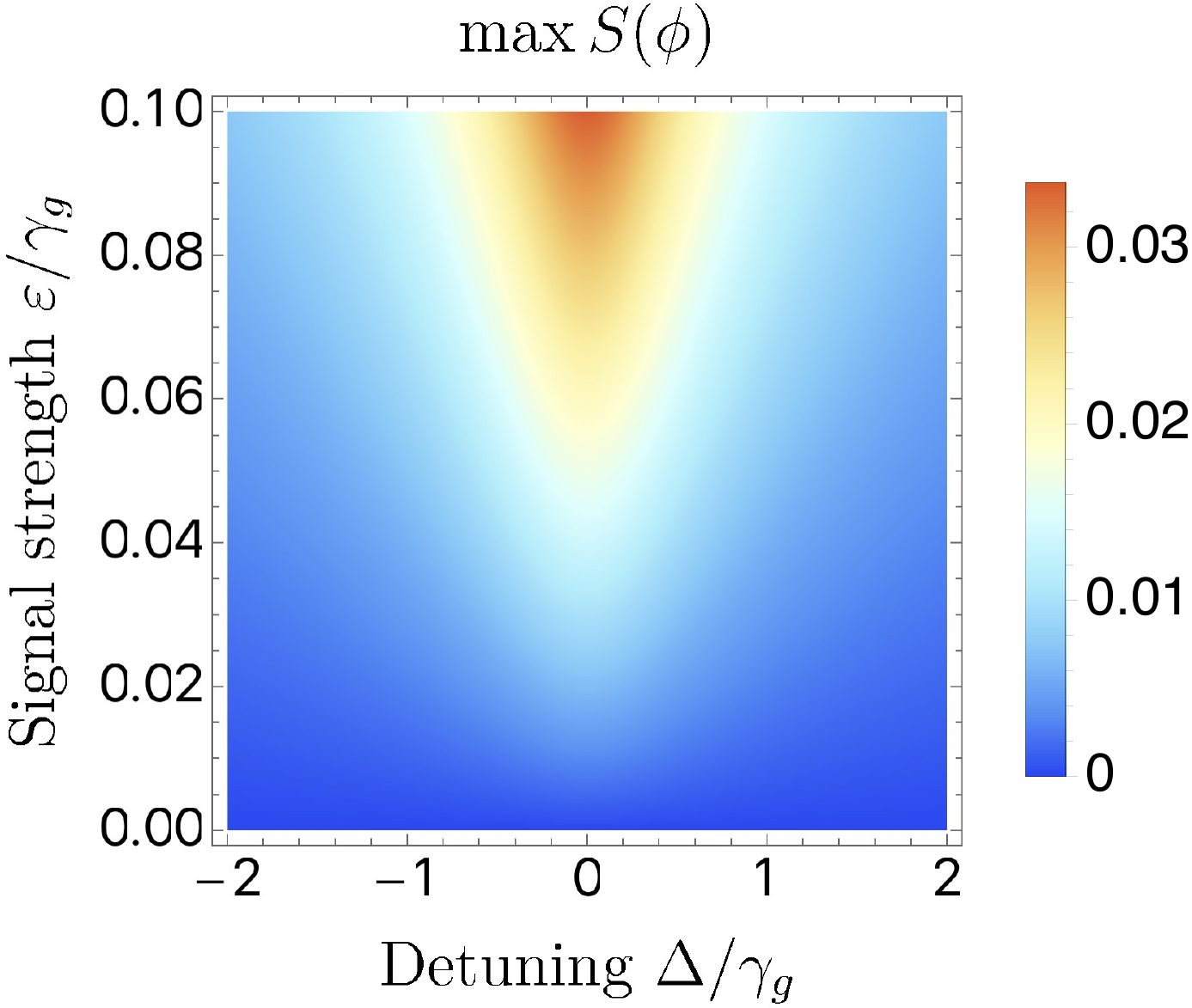}
		\put (0,84) {(c)}
	\end{overpic}
\caption{\label{fig:PL} The distribution of the phase $S(\phi)$ as a
  measure of phase locking for $\gamma_g=0.1\gamma_d$. (a) The phase locks to different values when the detuning $\Delta$ of the signal is varied. This is indicated by the dashed line which follows the peak $\max S(\phi)$ of the distribution. (b) Increasing the signal strength $\varepsilon$ leads to a more pronounced phase locking. (c) The Arnold tongue. Significant phase-locking is achieved over a broader range of detuning $\Delta$ by increasing the signal strength $\varepsilon$. The tongue does not touch the bottom axis due to the presence of noise.}
\end{figure*}

This behavior is illustrated in Fig.\,\ref{fig:PL}(a), where we show
how the distribution $S(\phi)$ is dragged along as the reference phase
is rotated by the detuning $\Delta$. Note that the amplitude of the
peak is also decreasing, as expected for a fixed signal strength
$\varepsilon$. To further characterize this dependence, we now fix the
detuning in Fig.\,\ref{fig:PL}(b), and find that $S(\phi)$ tends to be
more and more localized as we increase the signal strength. This is
characteristic of synchronization in the presence of noise, where a
stronger signal is able to reduce the tendency of
phase slips induced by the fluctuations of the phase. This can be seen
in Eq.\,\eqref{eq:Sdot}, which predicts that the synchronization
dynamics is characterized by a rate of order $\varepsilon$. As the signal
strength approaches $\gamma_\text{min}$, we thus expect the phase locking to be more and more pronounced. These features can be nicely summarized by the classic Arnold tongue, shown in Fig.\,\ref{fig:PL}(c). 

\begin{figure}
	\centering
	\includegraphics[width=\columnwidth]{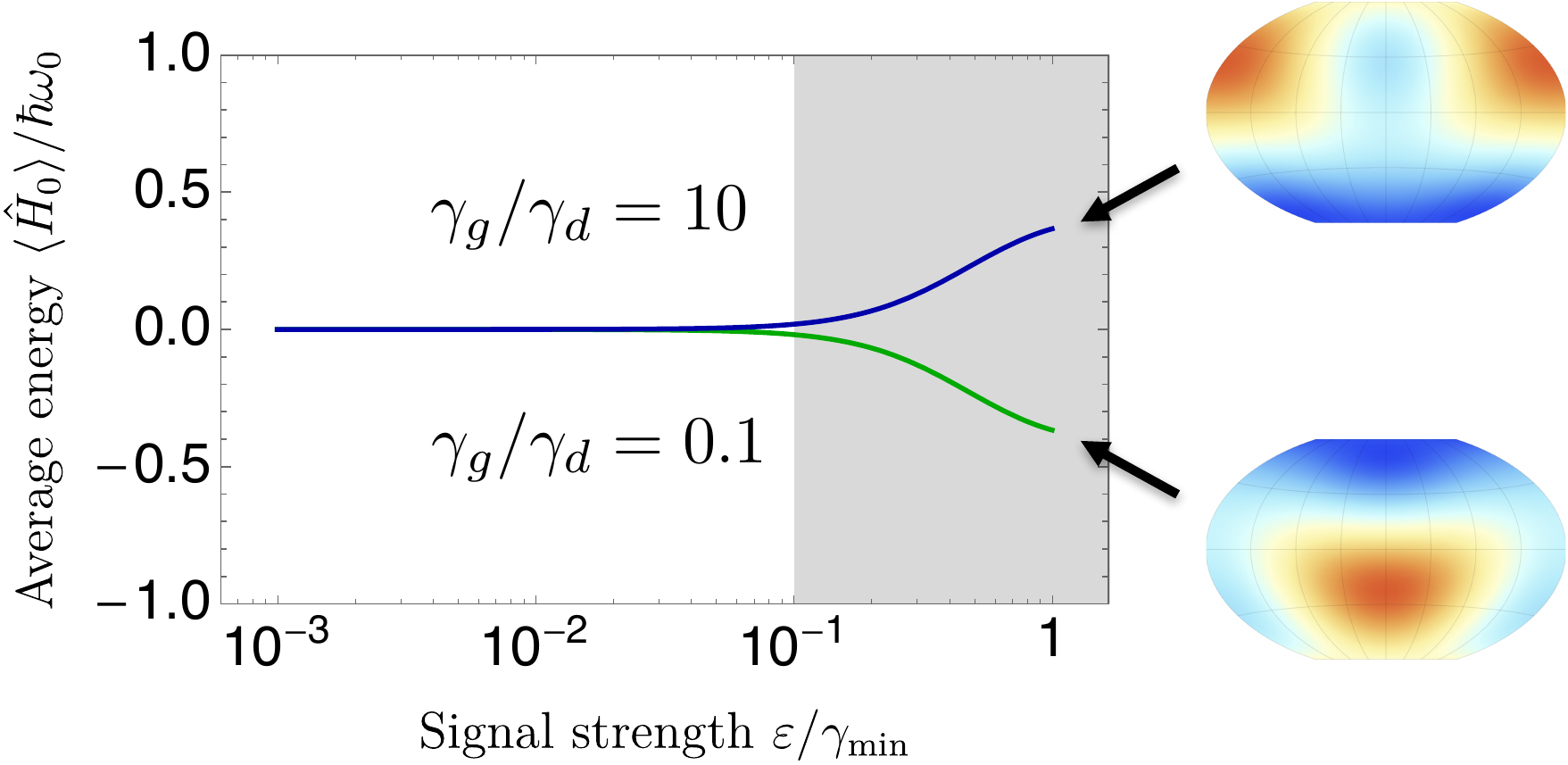}
\caption{\label{fig:breakLC} Influence of the signal strength
  $\varepsilon$ onto the limit cycle for zero detuning, $\Delta=0$. In the grey-shaded area, the effect of the signal is not limited to a perturbation of the free phase $\phi$ but instead substantially deforms the limit cycle by increasing or decreasing the energy of the spin, depending on the ratio $\gamma_g/\gamma_d$. The corresponding Q functions are shown for $\varepsilon=\gamma_\text{min}$.}
\end{figure}

Figures\,\ref{fig:PL}(b) and (c) appear to indicate that increasing
the signal strength $\varepsilon$ can only improve the resulting
phase locking. Yet, we expect a breakdown of this scaling from the
standard theory of synchronization~\cite{pikovsky01}: on increasing the signal
beyond the weakly-perturbed regime, the energy of
the self-oscillator starts to be affected and the stability of the limit
cycle is compromised. This is shown for $\Delta=0$ in Fig.\,\ref{fig:breakLC}, where we find that as the strength enters the gray region $\varepsilon> 0.1\gamma_\text{min}$, the energy of the spin is not stabilized anymore and the Q function starts to move away from the equator, \emph{i.e.} from its natural limit cycle.
\emph{Discussion.--} The framework developed here provides a natural
platform to explore how synchronization scales with the dimension of
the Hilbert space. In particular, one may conjecture an improvement of
the phase locking as the spin approaches the classical limit and
becomes less susceptible to quantum noise. To investigate this
direction, we apply the master equation~\eqref{eq:masterSynch} to the
case of a spin 2. We find that the spin synchronizes but with a weaker
phase locking: $\max S(\phi)=S(0)\approx 0.001$ at resonance for
$\varepsilon=0.1\gamma_g$ and $\gamma_g=0.1\gamma_d$. This is to be
compared with the value obtained for a spin 1 with the same
parameters: $\max S(\phi)=S(0)\approx 0.032$. This result highlights a
key open question in the field of quantum synchronization: given a
target limit cycle, what is the optimal dissipative map that will
yield the strongest phase locking once a signal is applied? For a generic spin number, the dynamics described by~\eqref{eq:masterDiss} is just one out of the many possibilities for stabilizing the limit cycle to the equator. For instance, an alternative choice in the high-spin limit would be to reproduce the Van der Pol structure of a linear gain and non-linear damping.

An advantage of the spin unit introduced here is that it can be implemented in a wide variety of physical systems. Specifically, any effective spin-1 ladder provides a suitable basis, such as that obtained in trapped ions~\cite{cohen14,senko15} or NV-centres~\cite{stark18}. The main requirement is the possibility to stabilize the target state $\ket{1,0}$, thereby providing the limit cycle to be synchronized. Incoherent gain and damping, for instance of the form of~\eqref{eq:masterDiss}, could be experimentally engineered by using the multi-level structure of these systems~\cite{poyatos96,lutkenhaus98}. The signal~\eqref{eq:Hsig} can then be implemented by a laser drive.
\emph{Conclusion.--} We have shown that the smallest possible system that can be synchronized is a single spin 1. We have demonstrated how the general theory of synchronization applies to this quantum system with no classical analogue. This was achieved by relying on the Husimi Q representation, which we propose as a powerful tool to study synchronization in spin systems.

The system presented here leads to many interesting research
directions. In particular, it can be used as an elementary building
block for studying synchronization of limit cycles in a network of quantum spins, minimizing the dimension of the resulting Hilbert space. The first step towards this direction would be a pair of spins, for which the interaction could be modeled by a direct effective Hamiltonian or a mediating bosonic field.
On a more fundamental level, studying how synchronization scales with the spin number leads to the open question of half-integer spins greater than 1/2 which do not have access to an eigenstate centered around the equator.
\begin{acknowledgments}
We would like to thank R. Fazio and N. L\"orch for discussions.
This work was financially supported by the Swiss SNF and the NCCR Quantum Science and Technology.
\end{acknowledgments}

\bibliography{spin1Bib}

\end{document}